\documentstyle[prl,aps,twocolumn,epsf]{revtex}
\newcommand{\lapprox}{\mbox{\raisebox{-4pt}{$\,\buildrel<\over\sim\,$}}}
\newcommand{\gapprox}{\mbox{\raisebox{-4pt}{$\,\buildrel>\over\sim\,$}}}
\begin{document}
\draft
\title{Crossover from Fermi liquid to Wigner molecule behavior in
quantum dots}
\author{R.~Egger,$^1$ W.~H\"ausler,$^1$ C.H.~Mak,$^2$ and H.~Grabert$^1$}
\address{${}^1$Fakult\"at f\"ur Physik, Albert-Ludwigs-Universit\"at,
 D-79104 Freiburg, Germany\\
${}^2$Department of Chemistry,
University of Southern California, Los Angeles, CA 90089-0482}
\date{Phys. Rev. Lett. 82, 3320 (1999), 83, 462(E) (1999)}
\maketitle
\begin{abstract}
The crossover from weak to strong correlations in parabolic quantum
dots at zero magnetic field is studied by numerically exact
path-integral Monte Carlo simulations for up to eight electrons.
By the use of a multilevel blocking algorithm, the simulations are 
carried out free of the fermion sign problem.
We obtain a universal crossover only governed by the  density parameter $r_s$.
For $r_s>r_c$,
the data are consistent with a Wigner molecule description,
while for $r_s<r_c$, Fermi liquid behavior is recovered.
The crossover value $r_c \approx 4$ is surprisingly small.
\end{abstract}

\narrowtext

Quantum dots can be considered as solid-state
artificial atoms with tunable properties.
Confining a small number of electrons $N$ in a two-dimensional
electron gas in semiconductor heterostructures, a number of
interesting effects arising from the interplay
between confinement and the Coulomb interaction between the electrons can  
be observed \cite{ashoori,leo}.
Since the confinement potential is usually quite shallow, the
long-ranged Coulomb interaction among the electrons
plays a prominent role, and in contrast to conventional
atoms effective single-particle approximations quickly
become unreliable. In the low-density (strong-interaction)
limit, $r_s\to \infty$, classical considerations
suggest a Wigner crystal-like phase
with electrons spatially arranged in shells \cite{classical}.
With quantum fluctuations, such a phase is best described
as a Wigner molecule.
In contrast, for high densities (weak interactions), $r_s\to 0$,
a Fermi liquid-like description is
expected to be valid, where it is more appropriate
to think of the behavior as resulting from
the single-particle orbitals being filled. The noninteracting
limit \cite{fock} is then typically used as a starting point for
the theoretical description of quantum dots.

To date, no reliable information exists for 
the crossover between these two limits.  This is mainly
due to a complete lack of sufficiently accurate methods
that are able to cover the full range of $r_s$, especially when no
magnetic field is present.
Exact diagonalization techniques are limited to very small
particle numbers and small $r_s$, otherwise a
large error due to the truncation of the Hilbert space arises \cite{merkt}.
Hartree-Fock calculations become increasingly unreliable
for large $r_s$ and are known to incorrectly 
favor spin-polarized states \cite{hf}.
Similarly, density functional calculations \cite{lda} introduce
uncontrolled approximations in the absence
of exact reference data. In principle, the quantum Monte Carlo (QMC) method
is the best candidate for producing reliable data for quantum dots.  
Unfortunately, the notorious fermion sign problem
makes direct QMC simulations almost impossible \cite{sign}. 
To avoid the sign problem,
the fixed-node approximation and a related variational approach
have been employed in Ref.~\cite{bolton}, but the results are no
longer exact.

In this study, we adopt a radically different approach to fermion 
QMC simulations, based on the recently developed multilevel
blocking (MLB) algorithm \cite{mlb,rtmlb}. The MLB algorithm is
able to provide numerically exact QMC results
free of the sign problem. In this Letter, we report large-scale
simulation results for the weak-to-strong-correlation crossover
for up to eight electrons. The numerical results
at large $r_s$ are shown to agree with a Wigner molecule description.

{\sl Model ---} We study a two-dimensional 
parabolic quantum dot at zero magnetic field,
\begin{equation}
H= \sum_{j=1}^N \left(\frac{\bbox{p}_j^2}{2m^*} +
\frac{m^*\omega_0^2}{2} \bbox{x}_j^2 \right) +
\sum_{i<j=1}^N \frac{e^2}{\kappa|\bbox{x}_i -\bbox{x}_j|} \;,
\end{equation}
where the positions (momenta) of the electrons are
denoted by $\bbox{x}_j \; (\bbox{p}_j)$. The effective mass is $m^*$,
and the dielectric constant is $\kappa$. 
The MLB calculations are carried out at fixed $N$
and fixed $S=(N_\uparrow- N_\downarrow)/2$, the $z$-component
of the total spin.
We present results for the energy,
$E = \langle H \rangle$, the radial charge and spin
densities $\rho(r)$ and $s_z(r)$ normalized to $\int_0^\infty dr \,
2\pi r \, \rho(r) = N$ and $\int dr \, 2\pi r \,
s_z(r) = S$, and the two-particle correlation function
\begin{equation} \label{cr}
g_S^{}(\bbox{x}) = \frac{2\pi l_0^2}{N(N-1)}
\left \langle \sum_{i\neq j=1}^N \delta(\bbox{x}-\bbox{x}_i+
\bbox{x}_j )\right \rangle \;.
\end{equation}
$g_S^{}$ is isotropic, and with 
$y=r/l_0$ prefactors are chosen such that $\int_0^\infty dy \,
y g^{}_S(y)=1$.  The length scale $l_0=\sqrt{\hbar/m^*\omega_0}$
from the confinement 
allows the interaction strength to be parametrized by $\lambda=l_0/a
=e^2/\kappa\omega_0 l_0$, where $a$
is the effective Bohr radius. For any given $N$ and $\lambda$,
the dimensionless density
parameter $r_s$ can then be obtained from the data, $r_s=r^*/a$,
where $r^*$ corresponds to the first maximum in $\sum_S g^{}_S(r)$.
The values for $r_s$ obtained this way agree well with the
predictions of an electrostatic point-charge model \cite{foot}.
In all simulations, the temperature was set to $T = 0.1\, \hbar
\omega_0/k_B$.

{\sl Method ---} The simulation method is based on a standard
discretized path-integral representation of the observables of interest,
where the sampling is done according to the MLB algorithm
described in detail in Ref.~\cite{mlb}.
A sample number \cite{mlb} of at most $K=600$ was sufficient to
eliminate bias from the algorithm,
and at the same time cured the sign problem.
The simulations have been carried out on up to five levels in the
MLB scheme. Data were collected from several $10^4$ samples
for each parameter set
$\{N,S,\lambda\}$, with a typical CPU time requirement of
a few days (for each set) on a SGI Octane workstation.
As a validation for this procedure, we have accurately reproduced the exact
diagonalization results for $N=2$ electrons \cite{merkt}.

\begin{figure}
\epsfxsize=1.0\columnwidth
\epsffile{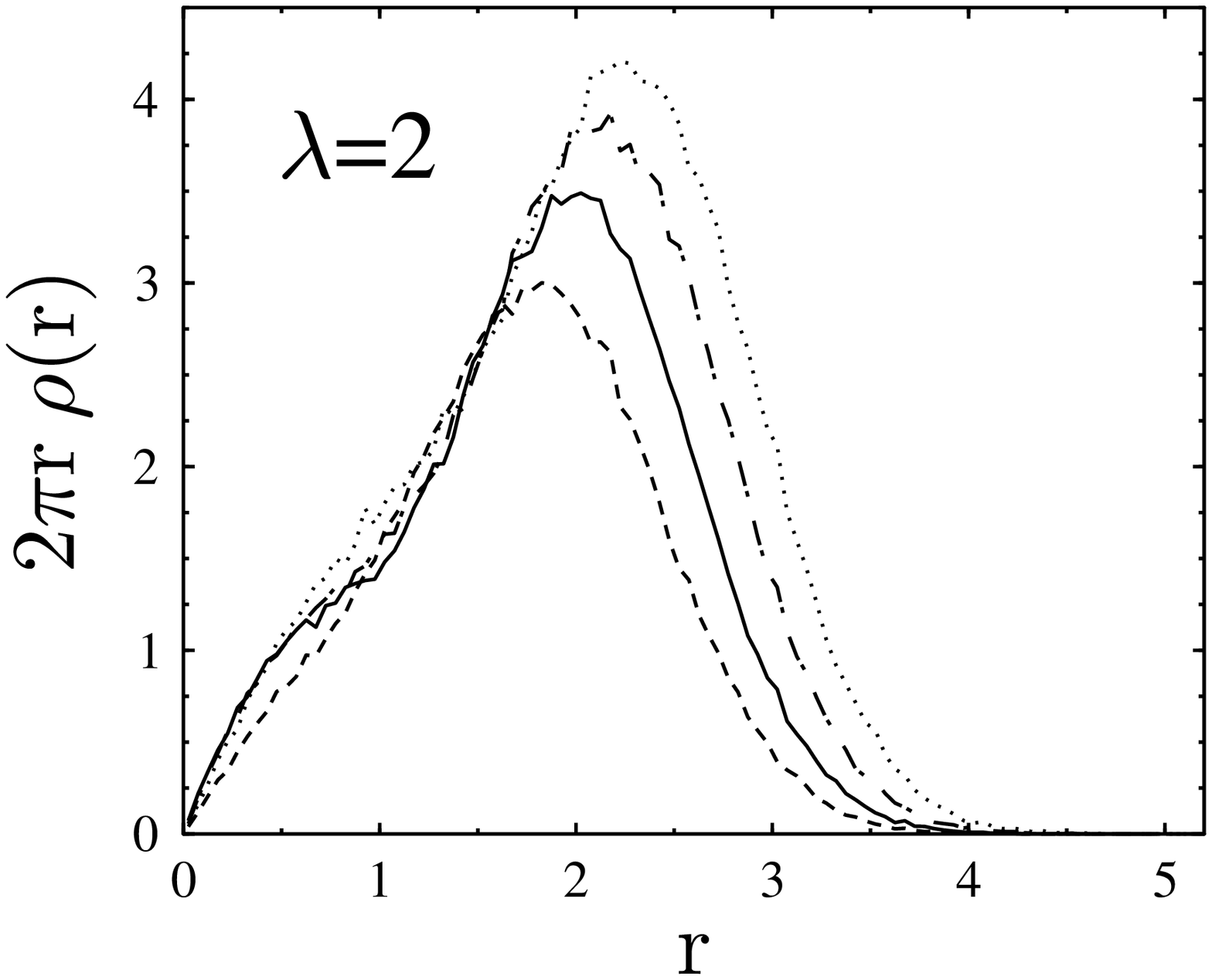}
\epsfxsize=1.0\columnwidth
\epsffile{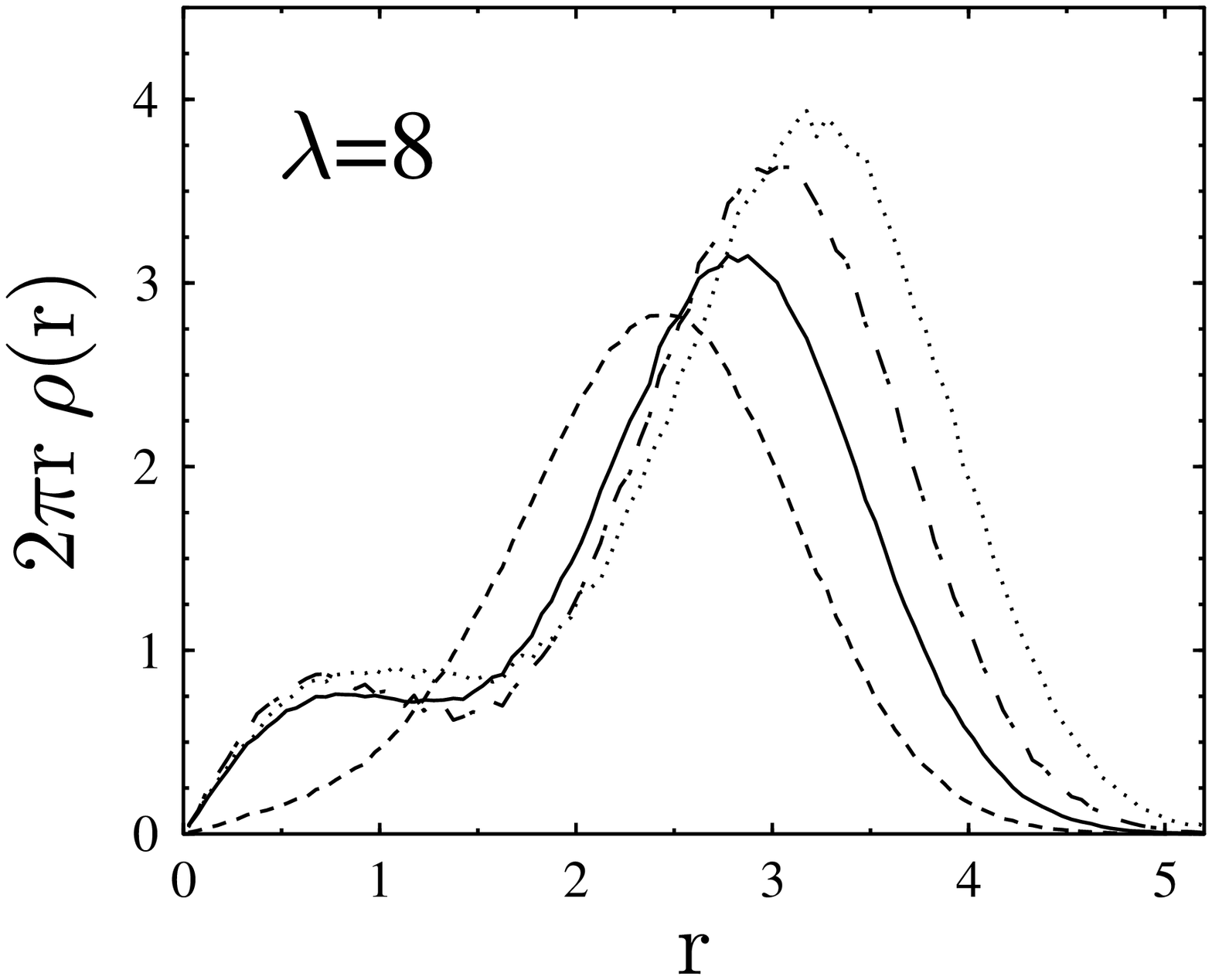}
\caption[]{\label{fig1}
Density $\rho(r)$ of the spin-polarized state ($S=N/2$)
for $\lambda=2$ and $\lambda=8$.
Dashed, solid, dashed-dotted, and dotted curves correspond to $N=5$,
6, 7, and 8, respectively. Units are such that $l_0=1$.
}
\end{figure}

{\sl Charge and spin densities ---}
Figure \ref{fig1} shows the charge density $\rho(r)$ of the
spin-polarized state for $N=5$ to 8 electrons. While for $\lambda=2$
increasing $N$ does not change $\rho(r)$ qualitatively,
the situation is different for strong interactions ($\lambda=8$),
signifying the onset of shell formation in real space. Such a
structure is clear evidence for Wigner molecule behavior.
The classical shell filling sequence has been predicted recently
\cite{classical}.
For $N<6$, there is only one shell, but the sixth electron
enters a new inner shell (1-5). Furthermore, for $N=7$ the
shell filling is 1-6, and for $N=8$ it is 1-7. These
predictions are in accordance with our data. Additional simulations
for up to 12 electrons at $\lambda=8$ [not shown here] further
verify the classical filling sequence.
The only exception is $N=10$, where we find a
3-7 instead of the predicted 2-8 structure.
Clear indications of a spatial shell structure at $N\geq 6$
can be observed even for
$\lambda = 4$, albeit significantly less pronounced than for $\lambda=8$.
For $\lambda \gapprox 4$, the charge densities are found to be
quite insensitive to $S$.  This is expected for a Wigner crystal where
particle statistics and spin influence energies or density
correlations only weakly. Our numerical results for the spin density
in this regime simply follow the corresponding charge density
according to $s_z(r)\simeq (S/N) \rho(r)$. 
A significant $S$-dependence of charge and
spin densities is observed only for weak correlations.

\begin{figure}
\epsfxsize=1.2\columnwidth
\epsffile{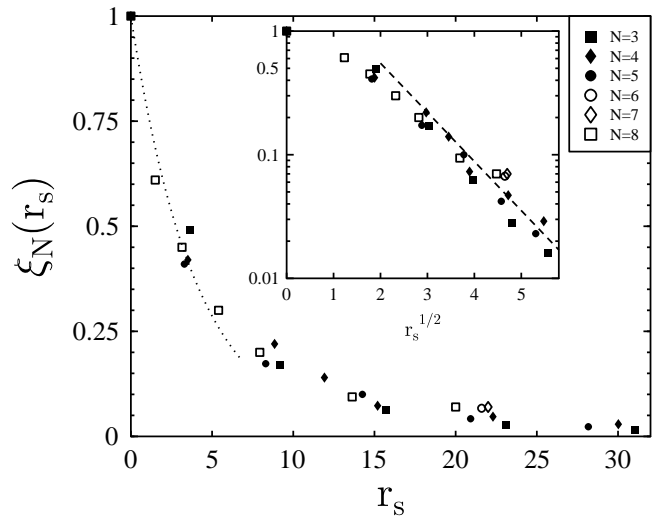}
\caption[]{\label{fig2} Numerical results for $\xi_N(r_s)$.
Statistical errors are of the order of the symbol size. The
dotted curve, given by $\exp(-r_s/r_c)$ with $r_c=4$, is a guide
to the eye only. The inset shows the same data on a
semi-logarithmic scale as a function of $\sqrt{r_s}$. The
dashed line is given by Eq.~(\ref{sc}).
}
\end{figure}

{\sl Crossover ---} To quantitatively investigate the
crossover from weak to strong correlations, we employ
the quantity
\begin{equation} \label{xi}
\xi_N(r_s) \propto
 \sum_{S,S'} \int_0^\infty dy \; y \, | g^{}_S(y)-g^{}_{S'}(y) | \;,
\end{equation}
normalized such that in the absence of interactions $\xi_N=1$.
The correlation function $g^{}_S(r)$ in Eq.~(\ref{cr})
is a very sensitive measure of Fermi statistics,
in particular revealing the spin-dependent correlation hole.  
Since interactions tend to destroy the Fermi surface,
the spin sensitivity of $g_S^{}(r)$ is largest for a Fermi gas, $r_s=0$.
In fact, for $r_s\to \infty$,  the 
correlation function $g_S(r)$ becomes completely spin-independent.
Hence the quantity $\xi_N(r_s)$ decays from unity at $r_s=0$ down to
zero as $r_s\to \infty$.  The functional form of this
decay is indicative of the crossover under consideration.

As seen in Figure \ref{fig2}, the crossover curve 
$\xi_N(r_s)$ becomes remarkably
{\sl universal}\, and depends only weakly on $N$. Its decay
defines a crossover scale $r_c$, where a simple exponential
fit for small $r_s$ yields $r_c\approx 4$.
For $r_s>r_c$, the functional form of $\xi(r_s)$ is better described by
\begin{equation}\label{sc}
\xi(r_s) \propto \exp\left(-\sqrt{r_s/r_c^\prime}\right) \;,
\end{equation}
where $r_c^\prime\approx 1.2$.   We mention in passing that 
the WKB estimate for $\langle \psi_S|H|\psi_S\rangle-
\langle\psi_{S'}| H |\psi_{S'}\rangle$ exhibits the same
large-$r_s$ dependence.  One can therefore argue that 
the spin sensitivity of the square of the
eigenfunctions $|\psi_S|^2$ also shows this behavior,
and thereby rationalize Eq.~(\ref{sc}).  
The value $r_c\approx 4$ is consistent with the appearance of
spatial shell structures in the density profile. In addition, 
the energy spectrum is in accordance with a
Wigner molecule description for $r_s>r_c$ (see below).
Summarizing, the crossover from weak to strong correlations
is characterized by the rather small value $r_c\approx 4$.

Parenthetically, we contrast this result with the values
$r_c^{\mbox{\tiny WC}}\approx 37$ found for clean
\cite{tanatar} and $r_c^{\mbox{\tiny dis}}\approx
7.5$ for disordered \cite{chui} unbounded systems.
The latter result provides evidence that breaking 
the continuous translation invariance stabilizes the crystallized
phase. The even smaller $r_c$ found here should then be due to
the confinement. Caution is adviced with the
thermodynamic limit,  $\omega_0\to 0$ with $r_s$ fixed, where
plasmon modes eventually govern the low-energy physics. There the value
$r_c^{\mbox{\tiny WC}}$ becomes relevant. For
GaAs-based structures with $r_s=4$, we estimate that 
spin-sensitive properties loose their significance only for
very large electron numbers, $N\gapprox 10^4$.

{\sl Energies ---} MLB results for the
energy at different parameter sets $\{N,S,\lambda\}$ are
listed  in Table \ref{table1}.  For given $N$ and $\lambda$, if the ground
state is (partially) spin-polarized with spin $S$, 
the simulations should yield the same energies for all $S'<S$. 
Within the accuracy of the calculation, this consistency check
is indeed fulfilled.

Detailed data are given in Table~\ref{table1}.
For $N=3$ electrons, as $r_s$ is increased, a transition occurs from the
$S=1/2$ to a spin-polarized $S=3/2$ ground state at an interaction strength
$\lambda\approx 5$ corresponding to $r_s\approx 8$.  
For $N=4$, we encounter a Hund's rule case.  By using perturbation
theory in $r_s$, one may show that the interactions lead to a
$S=1$ ground state. From our data, this standard Hund's rule (which
applies for small $r_s$) is seen to hold throughout the full range of $r_s$,
and the ground state spin stays $S=1$ even for large $r_s$.
A similar situation arises for $N=5$ electrons, where 
the ground state is characterized by $S=1/2$ for all $r_s$.
Turning to $N=6$, while one has filled orbitals and
hence a zero-spin ground state for weak correlations,
our results for $\lambda=8$ reveal a transition  to
a $S=1$ ground state as $r_s$ is increased.  A similar transition
from a $S=1/2$ ground state for weak correlations
to a partially spin-polarized $S=5/2$  ground state is 
found for $N=7$ electrons.  Finally, for $N=8$, 
as expected from  Hund's rule,
a $S=1$ ground state is observed for small $r_s$. 
 However, for $\lambda \gapprox 4$, corresponding
to $r_s\gapprox 10$, the ground state is seen to have spin $S=2$,
in contrast to the conventional Hund's rule prediction.

For strong correlations, $r_s>r_c$, the energy
levels and their spin splittings differ
considerably from what is expected from a Fermi liquid-like orbital picture.
In particular, the ground-state spin $S$ can change and
the excitation energy of higher-spin states becomes much
smaller than $\hbar\omega_0$.  This level structure reflects
the onset of shell formation in real space.
In fact, our large-$r_s$ data in Table \ref{table1}
 can be rationalized by starting from
crystallized electrons located at positions
fixed by electrostatics, and then evaluating the quantum corrections
due to particle exchange processes, namely rotation
and tunneling. Generalizing the method of Ref.\cite{zpb} by using
semiclassical estimates and group theory to satisfy the Pauli
principle, detailed predictions for low-energy spectra can be made
in terms of such a {\sl Wigner molecule}.  
While a detailed discussion of the Wigner molecule 
will be given elsewhere, it is
already apparent from our discussion above that
the value for $r_s$ where the ground state
spin changes is not given by $r_c\approx 4$ but is typically 
larger.  Therefore such transitions should be amenable to 
the Wigner molecule concept, which is indeed the case.

The above findings for the energy imply several 
novel and nontrivial consequences for transport experiments 
made by weakly coupling the dot to electrodes.
In particular, the addition energies following from Table~\ref{table1}
determine the positions of the conductance peaks measured experimentally
by capacitance spectroscopy \cite{ashoori} or by linear transport 
\cite{leo}.  Furthermore, high spins occurring close to
the ground-state energy can lead to negative
differential conductances in transport measurements, or even to
the disappearance of a conductance peak at low temperatures (spin blockade)
\cite{wein}. According to Table~\ref{table1}, 
$N=6$ would be a possible candidate in which to find 
negative differential conductances for the transition to
$N=5$ for $r_s\gapprox 15$, corresponding to 
$\hbar\omega_0\lapprox 0.4$~meV in a GaAs-based quantum dot. 
Furthermore, the spin-polarized ground state 
for $N=3$ at $\lambda\gapprox 5$ implies that the direct transition
into the $S=0$ ground state for $N=2$ is spin-forbidden. The
corresponding conductance peak should then disappear for
$\hbar\omega_0\lapprox 0.5$~meV. 
A similar situation arises for the $N=7$ to $N=6$ transition at
$\hbar \omega_0\lapprox 0.4$~meV.
Such phenomena cannot occur
in the weakly interacting regime $r_s<r_c$, where entering or escaping
electrons are accommodated in effective single-particle orbitals
together with their spins.   We note that sufficiently large
quantum dots allowing for experimental studies of the
Wigner molecule phase are within reach of current technology \cite{ash2}.

To conclude, we have presented numerically exact QMC results for
parabolic quantum dots covering the full crossover from weak
$(r_s\to 0)$ to strong ($r_s\to \infty$) correlations. 
The turnover from Fermi liquid to Wigner molecule like
behavior is basically independent of the particle number
and characterized by an astonishingly small crossover scale, $r_c\approx 4$.
Energy spectra in the low-density regime $r_s>r_c$ differ from 
single-particle
expectations but can be described within a Wigner molecule approach. 
Detailed predictions have been made for this Wigner molecule phase,
which should be directly accessible to current experiments.
It is straightforward
(and left to future MLB studies) to
study other confinements or interaction potentials, or to include
a magnetic field.

We thank M.~Wagner for providing the data of Ref.~\cite{merkt},
and R.~Bl\"umel, C.~Creffield, J.~Jefferson,
 and B.~Reusch for useful discussions.
W.~H.~acknowledges the hospitality of the Department of Physics at the
University of
Jyv\"askyl\"a, where parts of this work have been carried out.
This research has been supported by the SFB 276 of the Deutsche
Forschungsgemeinschaft (Bonn), by the Deutsche Akademischer
Austauschdienst, by the National Science Foundation
under grants CHE-9257094 and CHE-9528121, by the Sloan Foundation,
and by the Dreyfus Foundation.

\begin{table}
\caption{\label{table1}
MLB data for the energy for various $\{ N, S, \lambda\}$
parameter sets. Bracketed numbers denote statistical errors.
}
\begin{tabular}{llll||llll}
$N$ & $S$ & $\lambda$ & $E/\hbar\omega_0$&$N$ & $S$ & $\lambda$ &
$E/\hbar\omega_0$\\ \hline
3 & 3/2 & 2 & 8.37(1) & 5 & 5/2 & 8 & 42.86(4) \\
3 & 1/2 & 2 & 8.16(3) &  5 & 3/2 & 8 & 42.82(2) \\
3 & 3/2 & 4 & 11.05(1) & 5 & 1/2 & 8 & 42.77(4) \\
3 & 1/2 & 4 & 11.05(2) & 5 & 5/2 & 10 & 48.79(2) \\
3 & 3/2 & 6 & 13.43(1) & 5 & 3/2 & 10 & 48.78(3) \\
3 & 3/2 & 8 & 15.59(1) & 5 & 1/2 & 10 & 48.76(2) \\
3 & 3/2 & 10 & 17.60(1) & 6 & 3 & 8 & 60.42(2) \\
4 & 2 & 2 & 14.30(5) & 6 & 1 & 8 & 60.37(2) \\
4 & 1 & 2 & 13.78(6) & 7 & 7/2 & 8 & 80.59(4) \\
4 & 2 & 4 & 19.42(1) & 7 & 5/2 & 8 & 80.45(4) \\
4 & 1 & 4 & 19.15(4) &  8 & 4 & 2 & 48.3(2) \\
4 & 2 & 6 & 23.790(12) & 8 & 3 & 2 & 47.4(3) \\
4 & 1 & 6 & 23.62(2) & 8 & 2 & 2 & 46.9(3) \\
4 & 2 & 8 & 27.823(11) & 8 & 1 & 2 & 46.5(2) \\
4 & 1 & 8 & 27.72(1) & 8 & 4 & 4 & 69.2(1) \\
4 & 2 & 10 & 31.538(12) & 8 & 3 & 4 & 68.5(2) \\
4 & 1 & 10 & 31.48(2) & 8 & 2 & 4 & 68.3(2) \\
5 & 5/2 & 2 & 21.29(6) & 8 & 4 & 6 & 86.92(6) \\
5 & 3/2 & 2 & 20.71(8) &  8 & 3 & 6 & 86.82(5) \\
5 & 1/2 & 2 & 20.30(8) & 8 & 2 & 6 & 86.74(4) \\
5 & 5/2 & 4 & 29.22(7) & 8 & 4  & 8  & 103.26(5)  \\
5 & 3/2 & 4 & 29.15(6) & 8 & 3  & 8  & 103.19(4)  \\
5 & 1/2 & 4 & 29.09(6) & 8 & 2 & 8 & 103.08(4)\\
5 & 5/2 & 6 & 36.44(3) &  &  &  &  \\
5 & 3/2 & 6 & 36.35(4) &  &  &  & \\
5 & 1/2 & 6 & 36.26(4) &  &  &  & 
\end{tabular}
\end{table}

\end{document}